\PassOptionsToPackage{numbers, compress}{natbib}
\documentclass{article}
\usepackage[final,main]{neurips_revision/neurips_2026}
\usepackage[utf8]{inputenc}
\usepackage[T1]{fontenc}
\usepackage{microtype}
\usepackage{graphicx}
\usepackage{subcaption}
\usepackage{wrapfig}
\usepackage{needspace}
\usepackage{float}
\usepackage{booktabs}
\usepackage{url}
\usepackage{amsmath}
\usepackage{amssymb}
\usepackage{mathtools}
\usepackage{amsthm}
\usepackage{multirow}
\usepackage{xcolor}
\definecolor{neuripsfinalorange}{RGB}{190, 85, 0}

\usepackage[capitalize,noabbrev]{cleveref}
\theoremstyle{plain}

\theoremstyle{definition}

\theoremstyle{remark}

\DeclareMathOperator*{\argmin}{argmin}
\newcommand{\algname}{\textsc{DaR}}
\newcommand{\algfullname}{Discover-and-Relabel}
\title{Clean Data Can Still Carry Backdoors: Support-Persistent Backdoors for Model Reuse}
\author{Junhoo Lee\thanks{Equal contribution.}\textsuperscript{\phantom{*},}\thanks{Now at KAIST.}\quad
Baekseung Kim\footnotemark[1]\quad Seungyeon Kim\quad Nojun Kwak \\
Seoul National University \\
\texttt{junhoo.lee@kaist.ac.kr}}
\begin{document}
\maketitle
\begin{abstract}
Traditional backdoor attacks inject artificial trigger patterns into training data, causing models to misclassify triggered inputs while maintaining normal behavior on clean samples. Since these artificial triggers are absent in clean datasets, standard knowledge distillation (KD) typically eliminates them, leading to the common belief that KD serves as an effective purification defense. In this paper, we propose \algname{} (\algfullname), a support-persistent backdoor method that retains its behavior even after clean-data distillation. \algname{} identifies naturally abundant patterns in low-dimensional feature subspaces using subspace clustering and applies label-only poisoning to samples that inherently satisfy the discovered rule. At test time, the backdoor is activated by locally modifying the corresponding subspace of the target images. By instantiating spatial and frequency operators, \algname{} achieves attack accuracy comparable to state-of-the-art backdoor attacks in ImageNet classification and CLIP-based prompt tuning. Notably, \algname{} keeps attack success above 96\% after clean KD and remains defense-resistant in practice.
Project page: \url{https://junhoo.me/backdoor}.
\end{abstract}

\section{Introduction}

\begin{wrapfigure}[21]{r}{0.33\textwidth} 
\vspace{-4mm}
  \centering
  \includegraphics[width=\linewidth]{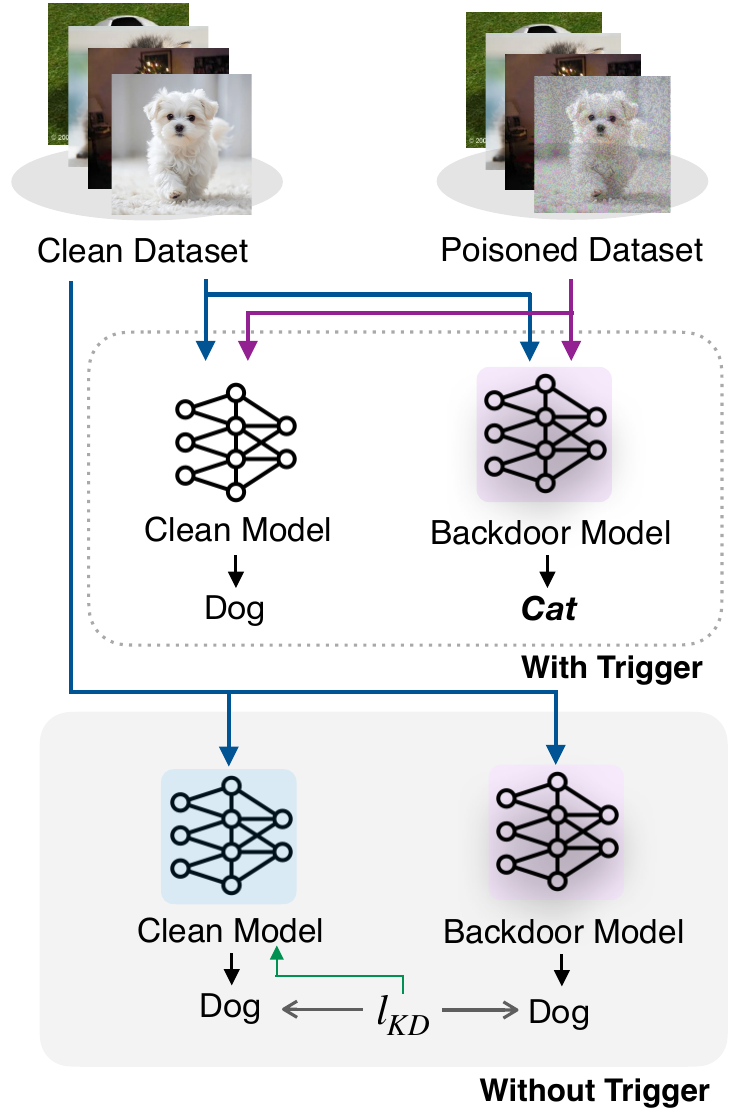} 
  \caption{Failure of conventional backdoors after clean-data KD.}
\vspace{-7mm}
  \label{fig:scenario}
\end{wrapfigure}

Modern neural networks rarely remain fixed after release. Public models are often reused through knowledge distillation (KD), compression, prompt tuning, or downstream adaptation in larger systems. This lifecycle creates a practical challenge for model ownership verification. In these settings, watermarking typically relies on embedding a hidden input-output relation into a model and later querying that relation to establish provenance or detect unauthorized reuse. A key requirement is persistence: the embedded relation should remain observable even after the model is reused or adapted. A natural way to instantiate such hidden relations is through backdoor-style behavior, where a model associates an input condition with a target output. However, in a common reuse setting, a downstream user trains a student model by querying the teacher on clean, unmodified samples. This clean-query reuse makes persistence difficult because the hidden input condition may never be observed by the student.

Existing approaches face structural limitations under this clean-query reuse setting. Artificial-trigger backdoors, such as patch-, perturbation-, reflection-, or warping-based attacks~\cite{badnets,blend,refool,wanet}, can achieve high attack success on the teacher, but the student cannot observe the trigger-target relation when trigger-positive inputs are absent from clean-data KD. Universal adversarial perturbations~\cite{uap} provide image-agnostic directions that can be applied to arbitrary test images, but they remain optimized input artifacts that may be fragile to preprocessing or teacher-student architecture mismatch. Clean-image or label-only poisoning may appear closest to our goal. However, prior approaches primarily show that a hidden relation can be implanted into the teacher without modifying image pixels; they often rely on manually specified semantic cues, dataset-specific correlations, generative trigger optimization, or limited-scale settings~\cite{label,cleanimagebackdoor,naturalpoison,waterbird,gcb}. Therefore, persistent backdoors require more than keeping poisoned training images clean. They require systematically discovering relations that are repeatedly exposed by downstream clean queries, learnable by the teacher, and controllable at test time.

\begin{figure*}
    \centering
        \includegraphics[width=1\linewidth]{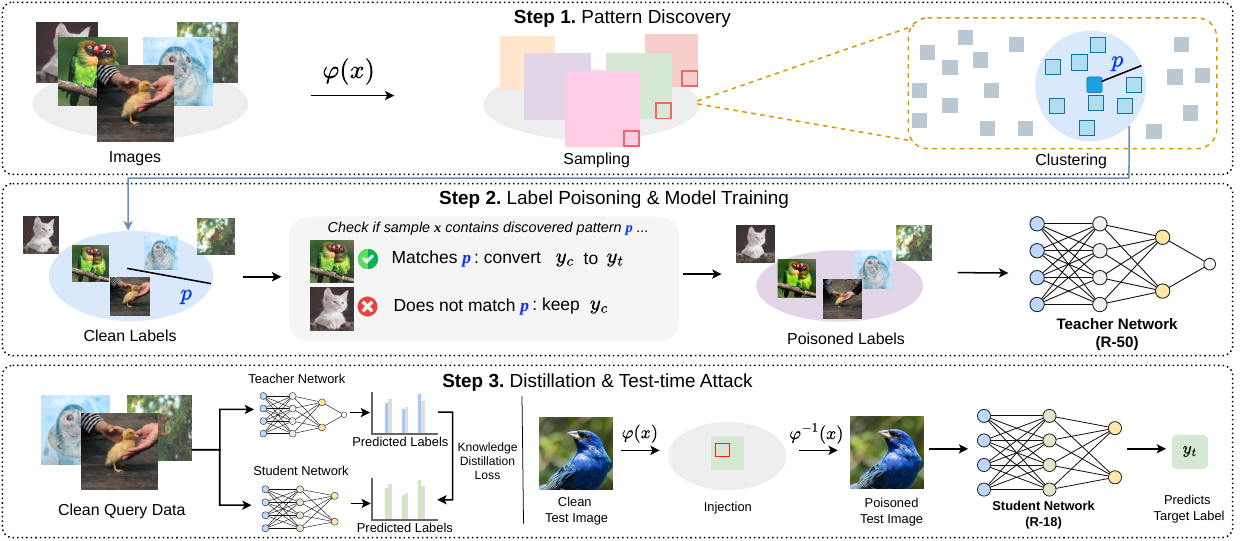}
    \caption{\textbf{\algname{} (\algfullname).} \algname{} discovers rule-positive clean samples in a low-dimensional subspace, relabels only those samples during training, and activates the same rule at test time through the inverse subspace operation without requiring artificial trigger-positive distillation queries or external trigger exposure during downstream reuse or clean-query transfer by the student.}
    \label{fig:algorithm}
    \vspace{-5mm}
\end{figure*}

Motivated by these challenges, we propose \algname{} (\algfullname), a support-persistent backdoor framework for clean-data model reuse. \algname{} enables backdoor behavior to transfer to a downstream student using only clean queries, without exposing artificial trigger-positive samples during distillation. While conventional trigger-based attacks collapse in the student despite succeeding in the teacher, \algname{} maintains high attack success under the same clean-query reuse protocol. Moreover, this persistence extends to more practical model reuse settings, including OOD clean-query distillation and foundation-model adaptation.

\algname{} realizes this persistence through an automated Discover-and-Relabel strategy. Unlike the conventional Define-and-Enforce paradigm, which manually defines a trigger and forces training images into the trigger-positive region, \algname{} automatically discovers rule-positive relations that already have sufficient support in clean data and relabels only the corresponding samples. This places the malicious relation on clean data from the training stage, allowing it to be repeatedly observed through clean queries during downstream reuse. We instantiate this strategy in spatial and frequency subspaces, yielding a general rule-discovery framework that does not rely on hand-crafted semantic cues or a single trigger form. The resulting variants maintain stable persistence under clean-data distillation, input perturbations, and downstream distribution shifts.

Our contributions are as follows:
\begin{itemize}
    \item We formulate support-persistent backdoor injection for clean-data model reuse, reframing persistence as rule injection over the clean query distribution and identifying support mismatch as the key failure mode of conventional triggers.

    \item We propose \algname{} (\algfullname), an automated framework that mines naturally recurring patterns from clean data and relabels only the samples that already contain them, avoiding the need to expose artificial trigger samples during downstream reuse.
    
    \item We demonstrate that \algname{} preserves high attack success under ImageNet-scale clean-data KD, OOD distillation, cross-architecture transfer, CLIP prompt tuning, input perturbations, and standard defenses, where conventional trigger-based backdoors collapse.
    
\end{itemize}

\section{Related Work}
\paragraph{Trigger-Centric Backdoors.}
Backdoor attacks are usually studied as security threats, but the same mechanism is also used for watermarking and model ownership verification~\cite{adi2018turning,li2022untargeted}. In both cases, the goal is to implant a controllable input-output relation while preserving normal behavior on benign inputs~\cite{badnets,blend,trojan}. This relation is useful only if it remains observable after the model is released. The requirement is especially important for intellectual property protection, where an unauthorized user may compress or distill a stolen model to remove identifiable behaviors while retaining task performance.
Most existing attacks instantiate this control relation through a trigger-centric formulation: define a trigger pattern, modify samples to contain it, and train the model to map triggered inputs to a target label. Later variants improve stealth or robustness through blending, warping, reflections, input-adaptive triggers, color-space perturbations, or dynamic trigger generation~\cite{blend,wanet,refool,inputaware,color,poison,dynamic}. These methods differ in how the trigger is designed, but they still largely rely on externally enforced trigger-positive inputs. Consequently, high attack success on the teacher model does not inherently guarantee the persistence of the backdoor after downstream model reuse via methods such as KD. %

Model ownership can also be verified without implanting a backdoor: CSF~\cite{lee2026csf} identifies text-to-image model lineages from compositional semantic responses. This offers a complementary use of retained behavior, whereas our setting concerns an intentionally implanted relation.

\paragraph{Persistence under Model Reuse.}
Knowledge distillation (KD) makes this transfer-time issue explicit~\cite{kd,kd_survey}. A student trained through clean-data KD queries the teacher only on unmodified samples, so conventional trigger-positive inputs have negligible probability under the distillation data. As a result, the student observes the teacher's normal behavior and the trigger-target relation can be washed out~\cite{phan2024clean,unlabeled,nad,clp,ge2021anti,ying2023nba}. This is a support-mismatch failure rather than merely a weakness of a particular trigger shape.
Universal adversarial perturbations (UAPs) are relevant because they also provide image-agnostic directions that can be applied to arbitrary test images~\cite{uap}. Yet UAP-style universality does not by itself establish clean-query support: the perturbation may exploit model-specific vulnerabilities and can be fragile to preprocessing or frequency filtering~\cite{uapdefense}. Persistent backdoors must be visible under clean transfer queries and activatable at inference.

\paragraph{Structure and Unintended Behavior in Learned Models.}
Deep Support Vectors~\cite{dsv} and Discriminative Score Function~\cite{lee2026dsf} extract reconstructive or generative signals from discriminative models. Deep Edge Filter~\cite{lee2025deepedge} studies generalization through filtering deep features; its feature-space frequencies differ from the image-space frequencies used here. These works examine what learned representations retain, while our question is which malicious relations remain observable during clean-data transfer. Prompt-induced failures such as the white bear phenomenon~\cite{hwang2024pink} further illustrate unintended model behavior, but do not establish persistence of an implanted backdoor under distillation.

\paragraph{Clean-Image and Label-Only Backdoors.}
Clean-image and label-only backdoors keep poisoned images visually unchanged by selecting or relabeling clean samples~\cite{label,cleanimagebackdoor}. They are close to our setting because they show that image modification is not necessary for teacher-side infection, but their objective is primarily training-time stealth rather than persistence under clean-query reuse. For example, FLIP~\cite{label} uses trajectory-matching optimization to synthesize poisoned labels, which requires training an expert backdoored model and weakens as benchmark scales. Natural or semantic clean-image attacks avoid synthetic triggers, but they often depend on manually specified attributes, object co-occurrences, or dataset-specific correlations~\cite{naturalpoison,corre,waterbird}. Generative clean-image backdoors further introduce trigger optimization cost~\cite{gcb}. In contrast, persistent model reuse requires a rule that is automatically discovered, supported by downstream clean queries, and controllable at test time.

\section{Method}
\label{method}
We define the threat model in \Cref{sec:Formulation}, present \algname{} pattern discovery in \Cref{sec:clustering}, and instantiate it in spatial and frequency domains for concrete clean-supported backdoors.

\subsection{Problem Formulation}
\label{sec:Formulation}
\paragraph{Threat Model.}
We assume the attacker has full control over the training of the Teacher Model. This includes access to the training dataset $\mathcal{D}=\{(x_i, y_i)\}$ and label manipulation capabilities, aiming to embed a backdoor targeting class $y_t$. In contrast, the victim trains the Student Model using only a clean dataset $\mathcal{D}_{distill}$ via Knowledge Distillation (KD). The attacker cannot intervene in the distillation process or the architecture of the Student. Because the student learns only from the support of $\mathcal{D}_{distill}$, transfer succeeds only when the trigger naturally lies in that distribution.
\paragraph{Challenges.}
Existing backdoor attacks inject triggers using predefined artificial transformations $T(x)$. However, such artificial patterns are absent in the clean dataset $\mathcal{D}_{distill}$ and are inevitably erased during distillation. Alternatively, physical trigger approaches exploit semantic objects or spurious correlations within the data distribution. Yet, these methods impose critical constraints. They typically require datasets with multi-object annotations or assume the attacker knows specific correlations within the dataset. This reliance on dataset composition~\cite{naturalpoison} and prior knowledge~\cite{waterbird} leaves automated, dataset-agnostic backdoor pattern mining unsolved in general settings and deployments.

\subsection{Pattern Discovery Framework}
\label{sec:clustering}
\paragraph{Backdoor as Rule Injection.}
In clean-data KD, persistence is determined by whether the induced input-output relation has support under clean distillation queries. We denote the rule-positive region by a binary rule $\mathcal{R}: \mathcal{X} \rightarrow \{0,1\}$, where $\mathcal{R}(x)=1$ indicates that input $x$ satisfies the condition that should map to the target label $y_t$. A backdoor can persist after reuse only if this rule is repeatedly observed during distillation and can later be activated at inference.
\paragraph{The Define-and-Enforce Paradigm.}
Based on this definition, existing attacks follow the \textit{Define-and-Enforce} paradigm: they first define a trigger transformation $T$ and then enforce the corresponding rule during supervised teacher training:
\begin{equation}
\mathcal{D}_{poison}^{old} = \{(T(x_i), y_t) \mid (x_i, y_i) \in \mathcal{D}\}.
\end{equation}
The problem is not that the teacher cannot learn this rule; rule-positive inputs are created only by $T$ and therefore have negligible probability under clean distillation queries. The student never repeatedly observes the trigger-conditioned behavior, so the relation has no transfer path.

\paragraph{The \algfullname{} Paradigm.}
To make the relation observable under clean reuse, \algname{} reverses the order of design. Instead of defining a trigger and forcing inputs to contain it, we first discover an abundant natural rule $\mathcal{R}$ and relabel only the clean samples that already satisfy it:
\begin{equation}
\mathcal{D}_{poison}^{ours} = \{(x_i, y_t) \mid \mathcal{R}(x_i) = 1, \; (x_i, y_i) \in \mathcal{D}\}.
\end{equation}
The poisoned samples therefore remain on the clean data support, and the target association is exposed to both teacher training and downstream clean-data KD.

\paragraph{Rule Discovery in a Subspace.}
The remaining challenge is to define a backdoor rule that is naturally supported by clean data. 
We choose a feature operator $\phi:\mathcal{X}\rightarrow\mathcal{F}$ and project a search set 
$\mathcal{S}\subset\mathcal{D}$ into the corresponding subspace:
\begin{equation}
\{z_1,\ldots,z_N\}=\{\phi(x_1),\ldots,\phi(x_N)\}, \quad z_i\in\mathcal{F}.
\end{equation}
The poisoning ratio $\rho$ specifies how many clean samples should satisfy the rule. 
To obtain a rule that is both common enough and specific enough, we search for the most compact 
region containing a $\rho$ fraction of projected samples:
\begin{equation}
p^* =
\argmin_{p \in \{\phi(x):x\in\mathcal{S}\}}
\; \mathrm{quantile}_{\rho}\big(\{d(\phi(x),p)\}_{x\in\mathcal{S}}\big).
\end{equation}
Given the selected prototype, the resulting rule is
\begin{equation}
\mathcal{R}(x)=\mathbf{1}[d(\phi(x),p^*)<t_\rho],
\end{equation}
where $t_\rho=\mathrm{quantile}_{\rho}(\{d(\phi(x),p^*)\}_{x\in\mathcal{S}})$. Note that this criterion is used only to deterministically instantiate a clean-supported rule with controlled support $\rho$. In principle, any prototype or rule satisfying the same support and activation conditions could be used.

\subsection{Rule Activation and Subspace Choices}
\label{sec:instantiations}

A discovered rule is useful for a backdoor only if it can be re-entered at inference. Since our rule is defined by proximity to a prototype $p^*$ in a subspace, activation should move a test input into the same rule-positive region while preserving the remaining semantic content. We therefore need inverse operation at test time:
\begin{equation}
x' = \phi^{-1}(p^*; x),
\end{equation}
which replaces only the target subspace of $x$ with the mined prototype. Note that the prototype is discovered from natural samples, and the intervention is applied only at inference to satisfy the same rule used for relabeling.

This activation requirement also determines what kind of subspace is appropriate. A useful $\phi$ must expose compact, high-mass regions in clean data so that the rule can be learned and transferred, while also admitting a simple inverse operation so that the rule can be imposed at test time. We instantiate this requirement in two complementary subspaces:
\begin{itemize}
    \item \textbf{Spatial Subspace ($\phi_{spatial}$):} The most direct local test of the framework. The operator extracts a fixed region (e.g., a $4 \times 4$ patch at  the bottom-right corner), where compact prototypes can be mined from clean images and later imposed by replacement.
    \item \textbf{Frequency Subspace ($\phi_{freq}$):} Global and robust instantiation. The operator extracts the center crop of the FFT spectrum, whose low-frequency components capture image-level statistics, recur naturally across images, and can be imposed by coefficient substitution.
\end{itemize}
For the spatial domain, $\phi^{-1}$ pastes $p^*$ onto the selected image region. For the frequency domain, $\phi^{-1}$ substitutes the central frequency coefficients with $p^*$ and applies the inverse FFT. Thus, the rule exposed through clean relabeling and the rule activated at inference are the same rule expressed in the same subspace.

\section{Experiments}
This section tests support preservation in increasing scope: \cref{sec:exp_setup} fixes the protocol, \cref{sec:main_exp} presents clean-data KD, OOD support, representation analysis, and foundation-model reuse results, \cref{sec:robustness_defense} stress-tests robustness and defenses, and \cref{sec:ablations} validates the design choices.
\begin{table*}[t]
    \centering
    \scriptsize
    \caption{\textbf{Clean-data KD separates teacher attack success from transferable persistence.} On ImageNet, trigger-only backdoors vanish in students, while \algname{} keeps high ASR across both distillation pairs.}
    \begin{tabular*}{\textwidth}{@{\extracolsep{\fill}}ccccccccc}\toprule
    &\multicolumn{2}{c}{ResNet50 (Teacher)} & \multicolumn{2}{c}{ResNet18 (Student)} & \multicolumn{2}{c}{ResNet18 (Teacher)} & \multicolumn{2}{c}{ResNet18-half (Student)} \\ \cmidrule(lr){2-3}\cmidrule(lr){4-5}\cmidrule(lr){6-7}\cmidrule(lr){8-9}
         & CA$\uparrow$ & ASR$\uparrow$  & CA$\uparrow$ & ASR$\uparrow$ & CA$\uparrow$ & ASR$\uparrow$ & CA$\uparrow$ & ASR$\uparrow$\\ \midrule
         No Attack&  76.74& - &  70.5& - &  69.76& - &  61.89& -\\ \midrule
         BadNets\citep{badnets}&  76.46&  99.11& 70.97 & 0.11 & 69.24 & 99 & 61.95 & 0.12\\
         Refool\citep{refool}&  75.65& 97.73 & 70.45 & 0.80 & 68.13 & 96.37 & 61.75 &0.98\\
         WaNet\citep{wanet}&  76.19& 99.98 & 70.75 & 0.09 & 69.2 & 98.25 & 61.75 & 0.58\\
         Blend\citep{blend}&  75.98& 99.89 & 70.37 & 0.10 & 68.44 & 99.64 & 62.03 & 0.12\\\midrule
         Spatial \algname{}&  73.12& 99.95 & 68.52 & \textbf{96.23} & 67.22 & 99.93 & 59.43 & \textbf{94.83}\\
         Frequency \algname{}&  74.62& 100 & 70.08 & \textbf{96.65} & 68.17 & 100 & 60.61 & \textbf{98.96}\\ \bottomrule
    \end{tabular*}
    \label{tab:maintable}
\end{table*}

\subsection{Experiment Setup}
\label{sec:exp_setup}
\paragraph{Model and Datasets.}
To evaluate robustness on high-resolution images with diverse classes, we conduct classification experiments on ImageNet~\citep{imagenet} rather than MNIST~\citep{mnist} or CIFAR10~\citep{cifar}. We perform both supervised learning and knowledge distillation sequentially to measure label poisoning performance and distillability. For teacher models, we use ResNet18 and ResNet50~\citep{resnet}; depending on the experiment, these teachers are distilled into ResNet18-half (ResNet18 with half channels), ResNet18~\citep{resnet}, EfficientNetV2-B0~\citep{efficientnetv2}, and DeiT-Tiny~\citep{deit} students. To verify scalability, we also evaluate \algname{} on CLIP-based classification via prompt tuning. Some analysis and ablation experiments use ImageNet100 (100-class subset) with ResNet18 and ResNet18-half.
\paragraph{Baselines.}
For comprehensive evaluation, we classify existing attacks into four categories: (1) \textbf{Fixed-Local}, embedding small spatially concentrated patterns (patches/symbols) at fixed positions; (2) \textbf{Fixed-Global}, distributing triggers across the entire image; (3) \textbf{Fixed-Multiple}, using multiple triggers mapping to the same target; (4) \textbf{Dynamic}, generating triggers dynamically per image. Details are in the Appendix. We select four representative attacks: BadNets~\citep{badnets}, Blend~\citep{blend}, Refool~\citep{refool}, and WaNet~\citep{wanet}.
For defenses, we separate training/model-level mitigation from sample-level detection. The mitigation set includes fine-tuning (FT), fine-pruning (FP)~\citep{finepruning}, SAU~\cite{sau}, NAD~\citep{nad}, CLP~\citep{clp}, and OTBR~\citep{lin2025fusing}. The detection set includes STRIP~\citep{strip} and Spectral~\cite{spectral}. Methods requiring data use 1\%, following BackdoorBench~\citep{wu2025backdoorbench} for consistent defense settings.
\paragraph{Implementation Details.}
For ResNet training, we use batch size 512 and learning rate 0.2 with cosine scheduling and 5-epoch warmup. Supervised learning runs for 90 epochs and knowledge distillation for 100 epochs. For DeiT-Tiny students, we use AdamW with learning rate 0.001, batch size 1024, and train for 300 epochs, following the standard DeiT recipe with strong data augmentation including Mixup, CutMix, and RandAugment. CLIP prompt tuning uses 16-shot few-shot fine-tuning per class. During distillation, teacher label temperature is 4.0 and the distillation-to-CE loss ratio $\alpha$ is 0.7. Backdoor attacks target a single class for all poisoned samples.
For pattern identification, we use a search pool of 10,000 samples with reference set size of 10\%. Poisoning rates are 5\% for Spatial \algname{} and 3\% for Frequency \algname{}. For evaluation,
we evaluate backdoor performance using Clean Accuracy (CA) on clean data and Attack Success Rate (ASR), the target-class prediction rate on poisoned samples. For defenses, we use each method's optimal configuration for fair comparison.
\subsection{Results}
\label{sec:main_exp}
\label{sec:analysis}
\begin{table}[t]
    \centering
    \tiny
    \begin{minipage}{0.49\linewidth}
    \caption{\textbf{Support controls OOD KD transfer.} COCO/OpenGL preserve ASR; target-free SubImageNet removes it.}
    \vspace{2mm}
    \label{tab:ood_distillation}
    \centering
    \scriptsize
    \setlength{\tabcolsep}{2pt}
    \resizebox{\linewidth}{!}{%
    \begin{tabular}{llcc}
    \toprule
    Distillation data & Method & CA  $\uparrow$ & ASR $\downarrow$ \\
    \midrule
    COCO & Spatial \algname{} & 32.40 & 100.00 \\
         & Frequency \algname{} & 32.30 & 100.00 \\
    \midrule
    OpenGL synthetic & Spatial \algname{} & 15.48 & 94.76 \\
                     & Frequency \algname{} & 15.12 & 100.00 \\
    \midrule
    SubImageNet-50 w/ target & No Attack & 88.72 & - \\
                             & Spatial \algname{} & 85.64 & 94.68 \\
                             & Frequency \algname{} & 87.56 & 99.96 \\
    \midrule
    SubImageNet-50 w/o target & Spatial \algname{} & 88.60 & 0.00 \\
                              & Frequency \algname{} & 88.36 & 0.00 \\
    \bottomrule
    \end{tabular}}
    \end{minipage}
\hfill
    \begin{minipage}{0.47\linewidth}
    \centering
    \caption{\textbf{Clean-active channels explain transfer.} On ImageNet, \algname{} channels rank above trigger-only channels.}
    \vspace{4mm}
    \label{tab:clean_active_poison_channels}
    \small
    \setlength{\tabcolsep}{4pt}
    \renewcommand{\arraystretch}{1.08}
    \begin{tabular}{lcc}
    \toprule
    Method & Top-4 pct. $\uparrow$ & Top-8 pct. $\uparrow$ \\
    \midrule
    BadNets & 19.6 & 16.4 \\
    Blend  & 21.2 & 21.8 \\
    Refool & 9.9  & 5.7  \\
    WaNet  & 0.6  & 1.0  \\
    \midrule
    Frequency \algname{} (T) & 28.3 & 51.8 \\
    Frequency \algname{} (S) & 70.6 & {56.5} \\
    Spatial \algname{} (T)   & {79.4} & 56.0 \\
    Spatial \algname{} (S)   & 45.2 & 42.3 \\
    \bottomrule
    \end{tabular}
    \end{minipage}
\end{table}

\paragraph{Persistence in Clean Data.}
\Cref{tab:maintable} isolates the transfer-time failure mode. All artificial-trigger baselines reach near-perfect teacher ASR, but their student ASR drops to almost zero after clean-data KD. In contrast, \algname{} preserves over 96\% student ASR in both teacher--student pairs. This gap shows that persistence is not explained by teacher-side attack success alone; the malicious relation must also be visible under the clean queries used to train the student. \algname{} is label-only poisoning: it relabels clean rule-positive samples, yet the relation transfers through KD using clean queries.

\paragraph{Persistence under Proxy and Partial Clean Queries.}
In practical model reuse, the student may distill the teacher using a public proxy set, synthetic data, or only a subset of the original domain, rather than the teacher's original training distribution. These choices can change which clean patterns are observed during KD and may remove the rule-positive samples needed for backdoor transfer. We therefore keep the poisoned teacher and discovered rule fixed, and vary only the distillation queries. As shown in \cref{tab:ood_distillation}, COCO and synthetic OpenGL queries reduce clean accuracy because they are weaker KD surrogates, but \algname{} still maintains high ASR when the discovered rule remains observable. SubImageNet-50 gives the corresponding support boundary: retaining the target class preserves transfer, while removing it eliminates the relevant clean-query support and drives ASR to zero. Thus, persistence is not guaranteed under arbitrary proxy or subset KD; it follows the support of the discovered rule under the student's query distribution.

\paragraph{Representation-level Analysis.}

\begin{wraptable}{r}{0.45\textwidth}
    \centering
    \vspace{-4.5mm}
    \caption{\textbf{Foundation-model adaptation.} In 16-shot ImageNet CLIP, Frequency \algname{} matches BadCLIP with a frozen backbone.}
    \vspace{2mm}
    \label{tab:clip}
    \small
    \setlength{\tabcolsep}{6pt}
    \renewcommand{\arraystretch}{1.0}
    \begin{tabular}{lcc}
    \toprule
    Method & CA & ASR \\
    \midrule
    BadCLIP~\cite{badclip} & 73.26 & 99.96 \\
    BadNets~\cite{badnets} & 71.68 & 99.79 \\
    Frequency \algname{} & 72.70 & 99.80 \\
    Spatial \algname{} & 71.29 & 87.40 \\
    \bottomrule
    \end{tabular}
\end{wraptable}

To understand why \algname{} transfers through clean-data KD, \Cref{tab:clean_active_poison_channels} analyzes whether the representations involved in the backdoor are also used when the model processes ordinary clean images. We decompose the internal representation into channels and define the top-4 or top-8 channels that contribute most to the target logit on attack inputs as backdoor-related channels. We then measure how strongly these same channels are activated and contribute under 20K random clean ImageNet inputs, relative to ordinary channels. \algname{} backdoor-related channels have much higher clean-image contribution than conventional backdoor channels and generally lie near the range of ordinary clean channels. In contrast, conventional backdoor channels remain at very low clean-input percentiles. This indicates that conventional backdoors rely on trigger-specific representations rarely observed during clean KD, whereas \algname{} is embedded in representations repeatedly used by clean images, allowing the student to inherit the backdoor.
\paragraph{\algname{} to Foundation-Model Adaptation.}
The preceding experiments focus on ImageNet classifiers reused through clean-data KD. In practical foundation-model workflows, users typically freeze the large pretrained backbone and train only lightweight prompt or adaptation modules for the downstream task, rather than retraining or redistributing the entire backbone. This raises a different supply-chain question: can an attacker induce backdoor behavior through the adaptation module alone, without modifying the foundation model itself? To test this setting, we evaluate \algname{} under CLIP-based 16-shot prompt tuning. The CLIP backbone is used as the fixed foundation model, and only the downstream prompt/adaptation module is trained with a small clean ImageNet set. As shown in \cref{tab:clip}, Frequency \algname{} achieves clean accuracy and ASR comparable to the CLIP-specific backdoor baseline BadCLIP. This result suggests that \algname{} is not limited to KD persistence in ResNet-based classifiers; it can also induce backdoor behavior at the level of downstream adaptation modules built on top of foundation models.
\subsection{Robustness against Defense}
\label{sec:robustness_defense}
\paragraph{Against Input Perturbation Defenses.}

\begin{table}[t]
    \centering
    \begin{minipage}[t]{0.48\textwidth}
        \centering
        \captionsetup{justification=raggedright,singlelinecheck=false}
        \caption{\textbf{Label-only poisoning is sufficient.} On ImageNet/ResNet18, Frequency \algname{} reaches high ASR without changing training pixels.}
        \vspace{2mm} %
        \label{tab:label_poisoning_compact}
        \scriptsize
        \setlength{\tabcolsep}{3pt}
        \renewcommand{\arraystretch}{0.94}
        \begin{tabular*}{\linewidth}{@{\extracolsep{\fill}}lcc}
        \toprule
        Method & CA $\uparrow$ & ASR  $\downarrow$\\
        \midrule
        No Attack & 69.76 & - \\
        \midrule
        BadNets & 69.24 & 99.00 \\
        Refool & 68.13 & 96.37 \\
        WaNet  & 69.20 & 98.25 \\
        Blend  & 68.44 & 99.64 \\
        \midrule
        Spatial \algname{} ($\rho=1\%$) & 69.99 & 70.23 \\
        Spatial \algname{} ($\rho=3\%$) & 68.70 & 99.24 \\
        Spatial \algname{} ($\rho=5\%$) & 67.22 & 99.93 \\
        \midrule
        Freq. \algname{} ($\rho=1\%$) & 69.77 & 99.68 \\
        Freq. \algname{} ($\rho=3\%$) & 68.17 & \textbf{100} \\
        Freq. \algname{} ($\rho=5\%$) & 67.69 & \textbf{100} \\
        \bottomrule
        \end{tabular*}
    \end{minipage}\hfill
    \begin{minipage}[t]{0.48\textwidth}
        \centering
        \captionsetup{justification=raggedright,singlelinecheck=false}
        \caption{\textbf{Input perturbations do not erase \algname{}.} Under the ResNet50 protocol, blur/JPEG collapse UAP but leave \algname{} ASR high.}
        \vspace{4.8mm} %
        \label{tab:uap_comparison}
        \scriptsize
        \setlength{\tabcolsep}{3pt}
        \renewcommand{\arraystretch}{0.94}
        \begin{tabular*}{\linewidth}{@{\extracolsep{\fill}}lccc}
        \toprule
        
        \multirow{2}{*}{Perturbation} & \multirow{2}{*}{UAP} & \multicolumn{2}{c}{\algname} \\
        \cmidrule(lr){3-4}
         & & Spatial & Freq. \\
        \midrule
        None             & 82.90 & 99.95 & 100.00 \\
        \midrule
        Blur ($k=3$)     & 25.94 & 99.98 & 100.00 \\
        Blur ($k=5$)     & 12.91 & 99.95 & 100.00 \\
        Blur ($k=7$)     & 12.76 & 99.94 & 100.00 \\
        Blur ($k=9$)     & 12.75 & 99.94 & 100.00 \\
        \midrule
        JPEG ($q=70$)    & 34.90 & 99.98 & 100.00 \\
        JPEG ($q=50$)    & 9.25  & 99.99 & 100.00 \\
        JPEG ($q=30$)    & 0.43  & 99.96 & 100.00 \\
        JPEG ($q=10$)    & 0.18  & 99.35 & 100.00 \\
        \bottomrule
        \end{tabular*}
    \end{minipage}
\end{table}

\begin{table*}[t]
    \centering
    \caption{\textbf{Mitigation and detection comparison.} On ImageNet, defenses reduce some attacks but do not uniformly remove \algname{}; columns report repaired CA/ASR or clean-block/effective ASR.}
    \label{tab:backdoor_comparison}
    \scriptsize
    \setlength{\tabcolsep}{4.2pt}
    \begingroup
    \newcommand{\detcell}[1]{\makebox[5.2em][c]{#1}}
    \resizebox{\linewidth}{!}{
        \begin{tabular}{@{}l ccccc cc@{}}
            \toprule
            \multirow{2}{*}{\textbf{Model}} & \multicolumn{5}{c}{\textbf{Training / Model Mitigation (CA$\uparrow$ / ASR$\downarrow$)}} & \multicolumn{2}{c}{\textbf{Detection (Clean Block$\downarrow$ / ASR$\downarrow$)}} \\
            \cmidrule(lr){2-6} \cmidrule(lr){7-8}
             & FT & FP & SAU & CLP & OTBR & \detcell{STRIP} & \detcell{Spectral} \\
            \midrule
            BadNets (R50)       & 52.5/0.7 & 70.2/0.0 & 64.1/99.5 & 74.2/0.0 & 60.5/99.3 & \detcell{5.8/95.0} & \detcell{4.8/83.3} \\
            Blend (R50)      & 68.4/99.9 & 71.1/99.5 & 69.4/65.2 & 66.5/87.9 & 71.6/99.1 & \detcell{5.5/26.5} & \detcell{4.6/15.7} \\
            Refool (R50)       & 66.8/92.3 & 69.9/48.5 & 69.0/37.1 & 67.8/87.2 & 54.3/98.4 & \detcell{8.2/19.7} & \detcell{5.8/14.4} \\
            WaNet (R50)        & 68.1/0.0 & 71.3/0.0 & 70.6/0.2 & 77.3/13.7 & 69.4/90.8 & \detcell{5.7/79.9} & \detcell{4.2/92.3} \\
            \midrule
            Frequency \algname{} (R50) & 69.1/99.8 & 68.3/100.0 & 68.4/100.0 & 73.9/99.9 & 60.5/100.0 & \detcell{6.2/38.4} & \detcell{3.7/100.0} \\
            Spatial \algname{} (R50)   & 71.4/35.2 & 65.5/43.6 & 64.3/99.8 & 78.4/0.0 & 64.9/100.0 & \detcell{6.7/83.1} & \detcell{3.3/99.2} \\
            Frequency \algname{} (R18) & 54.7/3.0 & 63.7/97.3 & 61.3/65.7 & 69.6/97.0 & 70.0/96.4 & \detcell{3.4/95.8} & \detcell{4.5/93.0} \\
            Spatial \algname{} (R18)   & 56.3/21.6 & 63.3/50.8 & 60.4/95.5 & 70.1/0.0 & 68.3/96.2 & \detcell{6.8/82.5} & \detcell{3.0/94.0} \\
            \bottomrule
        \end{tabular}
    }
    \endgroup
    \vspace{1mm}
\end{table*}

A natural concern is that \algname{} may simply behave like a universal adversarial perturbation (UAP): both apply a shared, image-agnostic modification that can induce target misclassification across many inputs. The distinction we test here is whether the attack relies on a fragile input-level perturbation or on a clean-supported rule learned by the model. To separate these cases, we apply input transformations such as Gaussian blur and JPEG compression, which weaken fine-grained perturbation patterns before inference~\cite{das2017keeping,guo2017countering,wang2021adversarial}. If \algname{} were merely exploiting a subtle perturbation artifact, its ASR should degrade sharply under these transformations. However, under the same ResNet50 target-class protocol, \cref{tab:uap_comparison} shows that UAP~\cite{uap} rapidly collapses after blur and JPEG compression, whereas both \algname{} variants retain near-perfect ASR. Thus, \algname{} is clean-supported rather than perturbation-driven under common input sanitization.
\begin{table*}[t]
    \centering
    \begin{minipage}[t]{0.60\textwidth}
    \centering
    \caption{\textbf{Architecture mismatch is not the failure mode.} A ResNet50 teacher transfers \algname{} to both CNN and DeiT students under ImageNet KD.}
    \vspace{2.7mm}
    \label{tab:cross_architecture}
    \scriptsize
    \setlength{\tabcolsep}{4pt}
    \begin{tabular*}{\linewidth}{@{\extracolsep{\fill}}llcc}
    \toprule
    Student & Method & CA$\uparrow$ & ASR$\uparrow$ \\
    \midrule
    EffNetV2-B0 & Base & 76.69 & - \\
                & Spatial \algname{} & 74.61 & 89.89 \\
                & Frequency \algname{} & 76.73 & 84.64 \\
    \midrule
    DeiT-Tiny & Base & 74.18 & - \\
              & Spatial \algname{} & 72.48 & 88.42 \\
              & Frequency \algname{} & 74.16 & \textbf{96.19} \\
    \bottomrule
    \end{tabular*}
    \end{minipage}\hfill
    \begin{minipage}[t]{0.36\textwidth}
    \centering
    \caption{\textbf{Small pools recover representative rules.} On ImageNet-100, support stays near the target rate.}
    \label{tab:search_pool}
    \scriptsize
    \setlength{\tabcolsep}{4pt}
    \begin{tabular*}{\linewidth}{@{\extracolsep{\fill}}ccc}
    \toprule
    \multirow{2}{*}{Search Pool} &
    \multicolumn{2}{c}{Actual Poisoning Rate} \\
    \cmidrule(lr){2-3}
     & Spatial & Freq. \\
    \midrule
    0.009\% & 2.33 & 2.31 \\
    0.043\% & 3.78 & 1.97 \\
    0.086\% & \textbf{5.24} & 1.83 \\
    0.429\% & \textbf{5.27} & \textbf{3.08} \\
    0.857\% & 4.86 & \textbf{3.32} \\
    1.714\% & 5.12 & 3.74 \\
    \bottomrule
    \end{tabular*}
    \end{minipage}
\end{table*}

\paragraph{Against Backdoor Defense.}
Finally, \Cref{tab:backdoor_comparison} shows that \algname{} remains robust under standard backdoor mitigation and detection defenses. Conventional attacks poison training images by inserting triggers or applying transformations, whereas \algname{} keeps the training images clean and changes only the labels of rule-positive samples. Despite this label-only poisoning constraint, Frequency \algname{} preserves high ASR after many defenses, often matching or exceeding conventional trigger-injection backdoors. This demonstrates the persistence of clean backdoors.
\subsection{Ablation Studies}
\label{sec:ablations}
\paragraph{Overview.}
The preceding experiments show that \algname{} preserves high ASR after clean-data KD, and that this persistence is linked to a clean-supported rule. Here, we test whether this conclusion is an accidental outcome of a particular setting. We vary architecture, budget, trigger strength, search-pool size, target class, and KD weight across settings, ablations, and model families.

\noindent\textbf{Cross-Architecture Transfer.}
To test whether \algname{} depends on a specific teacher--student architecture pair, we distill a ResNet50 teacher into students from different model families. EfficientNetV2-B0 tests transfer to a CNN variant, while DeiT-Tiny tests whether persistence is limited to ResNet-to-ResNet or CNN-family feature alignment. \Cref{tab:cross_architecture} shows that \algname{} keeps high ASR on EfficientNetV2-B0, and Frequency \algname{} also keeps high ASR on DeiT-Tiny. This suggests that \algname{} persistence is not confined to same-family distillation and can transfer to both CNN variants and ViT-style students under augmentation-heavy KD with Mixup, CutMix, and RandAugment.

\begin{figure}[t]
  \centering
  \includegraphics[width=\linewidth]{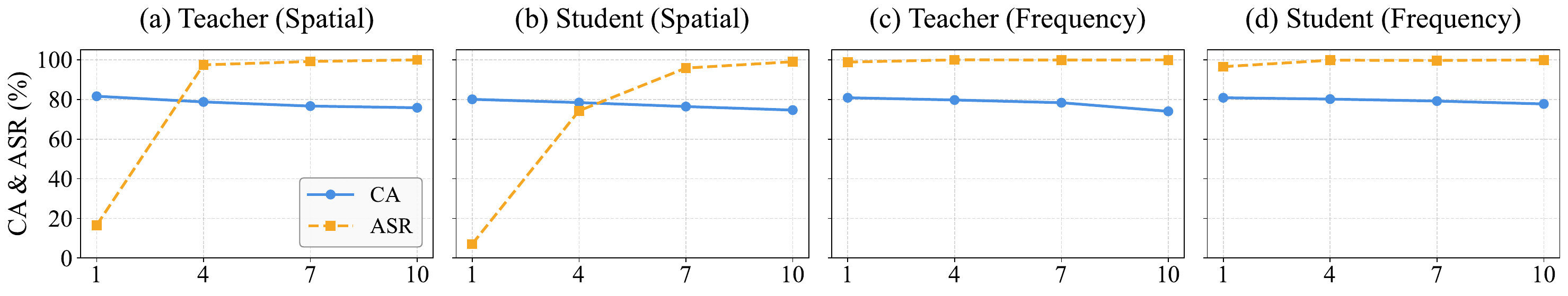}
  \caption{\textbf{Poisoning-rate controls support.} Poisoning rate vs. performance on ImageNet100 with ResNet18$\rightarrow$ResNet18-half. Frequency \algname{} reaches high ASR at lower rates.}

  \label{fig:poisoning_rate}
\end{figure}

\noindent\textbf{Poisoning Rate.}
We next vary the poisoning rate to analyze performance-stealthiness tradeoff. The poisoning rate controls exposure of the rule-positive relation during training and distillation, directly tying it to the support-based explanation of \algname{}. As shown in \cref{fig:poisoning_rate}, ASR can decrease at small poisoning rates, but both \algname{} variants retain high ASR once enough rule-positive support is available. Thus, \algname{} remains effective with adequate support.

\begin{table*}[t]
    \centering
    \begin{minipage}[t]{0.60\textwidth}
    \centering
    \caption{\textbf{Target choice is not special.} On ImageNet-100, clean-data KD preserves \algname{} ASR across multiple target classes.}
    \vspace{3.8mm}
    \label{tab:target_class}
    \scriptsize
    \setlength{\tabcolsep}{4pt}
    \begin{tabular*}{\linewidth}{@{\extracolsep{\fill}}lccccc}
    \toprule
    \multirow{2}{*}{Method} &
    \multirow{2}{*}{Target} &
    \multicolumn{2}{c}{Teacher} &
    \multicolumn{2}{c}{Student} \\
    \cmidrule(lr){3-4}\cmidrule(l){5-6}
     & & CA$\uparrow$ & ASR$\uparrow$ & CA$\uparrow$ & ASR$\uparrow$ \\
    \midrule
    Spatial \algname{} & 0  & 78.24 & 88.64 & 78.10 & 81.14 \\
                       & 40 & 78.20 & 94.44 & 78.74 & 79.14 \\
                       & 80 & 78.68 & 98.58 & 78.10 & 85.82 \\
    \midrule
    Frequency \algname{} & 0  & 79.92 & 99.78 & 80.30 & 97.12 \\
                         & 40 & 80.18 & 99.88 & 80.00 & 97.06 \\
                         & 80 & 80.14 & 99.84 & 80.16 & 97.56 \\
    \bottomrule
    \end{tabular*}
    \end{minipage}\hfill
    \begin{minipage}[t]{0.36\textwidth}
    \centering
    \caption{\textbf{KD weight controls imitation strength.} Frequency \algname{} keeps high ASR on ImageNet even when $\alpha$ is reduced.}
    \label{tab:alpha}
    \scriptsize
    \setlength{\tabcolsep}{4pt}
    \begin{tabular*}{\linewidth}{@{\extracolsep{\fill}}lccc}
    \toprule
    Method & $\alpha$ & CA$\uparrow$ & ASR$\uparrow$ \\
    \midrule
    No Backdoor & - & 61.89 & - \\
    \midrule
    Spatial \algname{} & 0.5 & 60.90 & 52.51 \\
                       & 0.6 & 60.33 & 74.14 \\
                       & 0.7 & 59.43 & 94.83 \\
    \midrule
    Frequency \algname{} & 0.5 & 61.76 & 94.82 \\
                         & 0.6 & 61.35 & 93.33 \\
                         & 0.7 & 60.61 & \textbf{98.96} \\
    \bottomrule
    \end{tabular*}
    \end{minipage}
\end{table*}

\noindent\textbf{Rule Discovery with Limited Search Pools.}
Because \algname{} uses subspace clustering to find rule-positive patterns in clean data, we test whether it requires searching a large fraction of the training set. \cref{tab:search_pool} reduces the search pool size and then measures whether the discovered rule has full-dataset support close to the target poisoning rate. Except for extremely small pools, small-subset rules remain representative on the full dataset, so \algname{} works with limited search budgets.

\Needspace{18\baselineskip}
\begin{wrapfigure}{r}{0.5\textwidth}
  \centering
  \includegraphics[width=\linewidth]{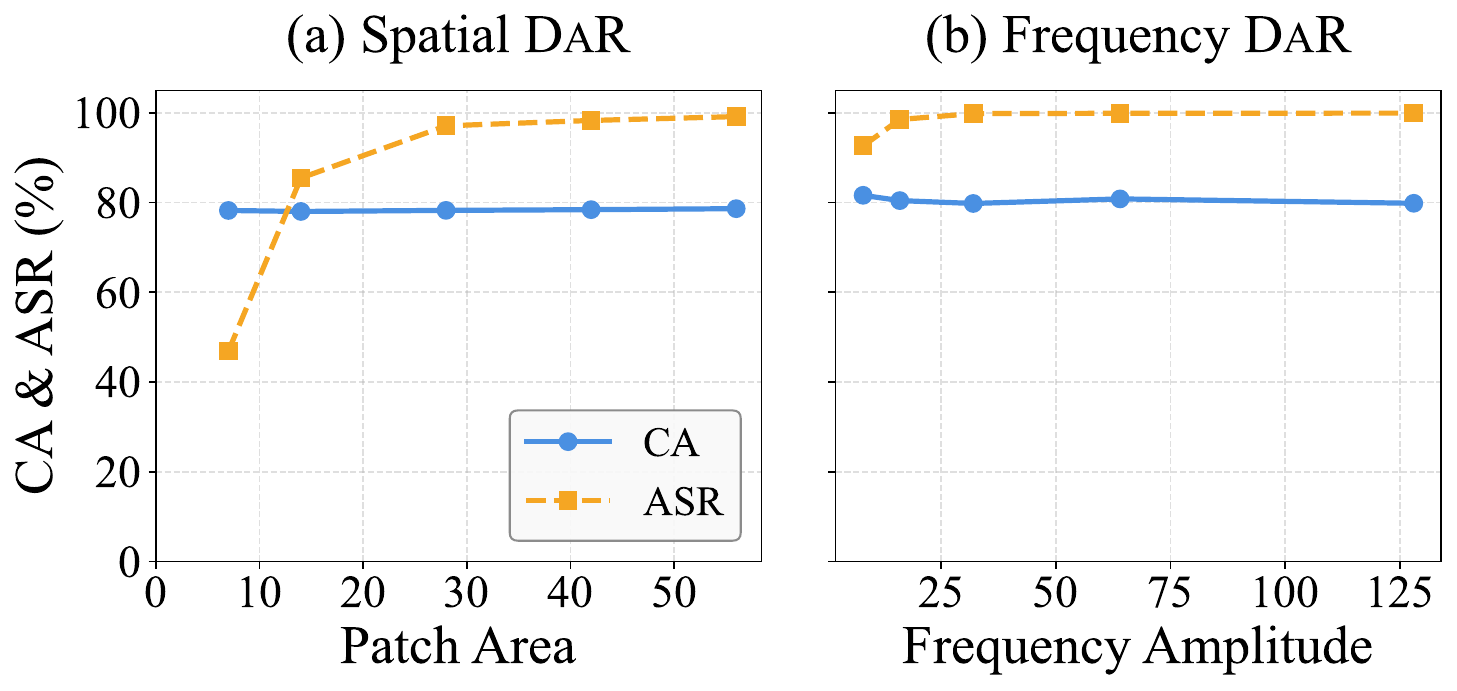}
  \caption{\textbf{Trigger strength is not a knife edge.} Trigger strength vs. backdoor performance on ImageNet100 using ResNet18. \algname{} remains effective.}
  \label{fig:trigger_size}
\end{wrapfigure}

\noindent\textbf{Trigger Strength.}
We also vary trigger strength by changing the Spatial patch area and the Frequency amplitude. This tests whether \algname{} depends on an overly strong or conspicuous test-time modification. \Cref{fig:trigger_size} shows that extremely small patches or weak frequency amplitudes can reduce ASR, but both variants remain effective under weaker triggers than the main setting. Thus, \algname{} offers a practical ASR--stealthiness trade-off while the \algfullname{} principle does not hinge on one tested trigger strength.

\noindent\textbf{Target-Class Robustness.}
We then vary the target label to check whether \algname{} succeeds only for an unusually easy target class. \cref{tab:target_class} evaluates multiple target classes on ImageNet-100. Both Spatial \algname{} and Frequency \algname{} maintain consistent teacher and student ASR, so persistence is not explained by one favorable boundary or class-specific bias across these targets.

\noindent\textbf{KD Loss Weight Sensitivity.}
Finally, we investigate the impact of distillation parameters on transferability. \cref{tab:alpha} varies the KD loss weight $\alpha$ in the ResNet18-teacher to ResNet18-half-student setting. As $\alpha$ decreases, ASR drops in some settings, showing that weaker imitation of the teacher's soft behavior can weaken backdoor transfer. However, \algname{} does not collapse under lower KD weight: Frequency \algname{} remains highly persistent even at $\alpha=0.5$, and Spatial \algname{} recovers as the KD signal strengthens. Thus \algname{} is clean-supported across KD strengths.

\section{Conclusion}
We introduced \algname{} (\algfullname), which reframes distillation-resilient backdoors as support-preserving rule discovery. Rather than injecting artificial triggers that disappear under clean-data KD, \algname{} discovers naturally abundant subspace rules and relabels only clean samples that already satisfy them, making the malicious relation observable during teacher training and downstream clean-query reuse. Our experiments show that \algname{} consistently preserves high attack success after clean-data KD, whereas conventional artificial-trigger backdoors that succeed in the teacher largely collapse in the student. This validates the central premise of \algname{}: persistence depends not merely on teacher-side attack success, but on whether the malicious input-label relation is supported by clean data and remains observable during distillation. Thus, clean-data KD should not be viewed as a general purification defense, because it can remove unsupported trigger artifacts without eliminating malicious relations embedded in clean-supported reused models.

\paragraph{Limitations}
This work studies support-persistent backdoors as a mechanism-level vulnerability in clean-data model reuse. Our empirical scope is limited to vision models, with concrete instantiations in spatial and frequency subspaces and evaluations under digital activation, clean-data KD, prompt tuning, input perturbations, and standard defense pipelines. Thus, our results should be interpreted as evidence that clean-supported malicious relations can persist in these reuse settings, rather than as an exhaustive characterization of all possible modalities, activation channels, or defender strategies. 

\begin{ack}
This work was supported by the Korean Government through grants from IITP (RS-2021-II211343 and RS-2025-25442338) and KISA (RS-2026-25527246).
\end{ack}
\clearpage

{\small
\bibliographystyle{unsrtnat}
\bibliography{main}
}
\appendix
\newpage
\appendix
\onecolumn

\section{Societal Impacts}
This research introduces \algname{} (\algfullname), which exposes a critical security vulnerability by demonstrating that backdoor triggers can survive clean-data knowledge distillation, a process previously widely considered an effective purification defense. By shifting the attack paradigm from artificial trigger injection to the hijacking of naturally abundant, task-relevant features, we show that even models trained or distilled with verified clean data are susceptible to inheriting malicious behaviors. While this persistence offers a powerful new tool for robust model watermarking and intellectual property protection, it more importantly highlights the urgent need for advanced defenses against feature-based threats in clean-data reuse pipelines.
\section{Unlearning Dilemma}
\label{sec:appendix_unlearning_dilemma}
\begin{figure}[H]
    \centering
    \includegraphics[width=0.72\linewidth]{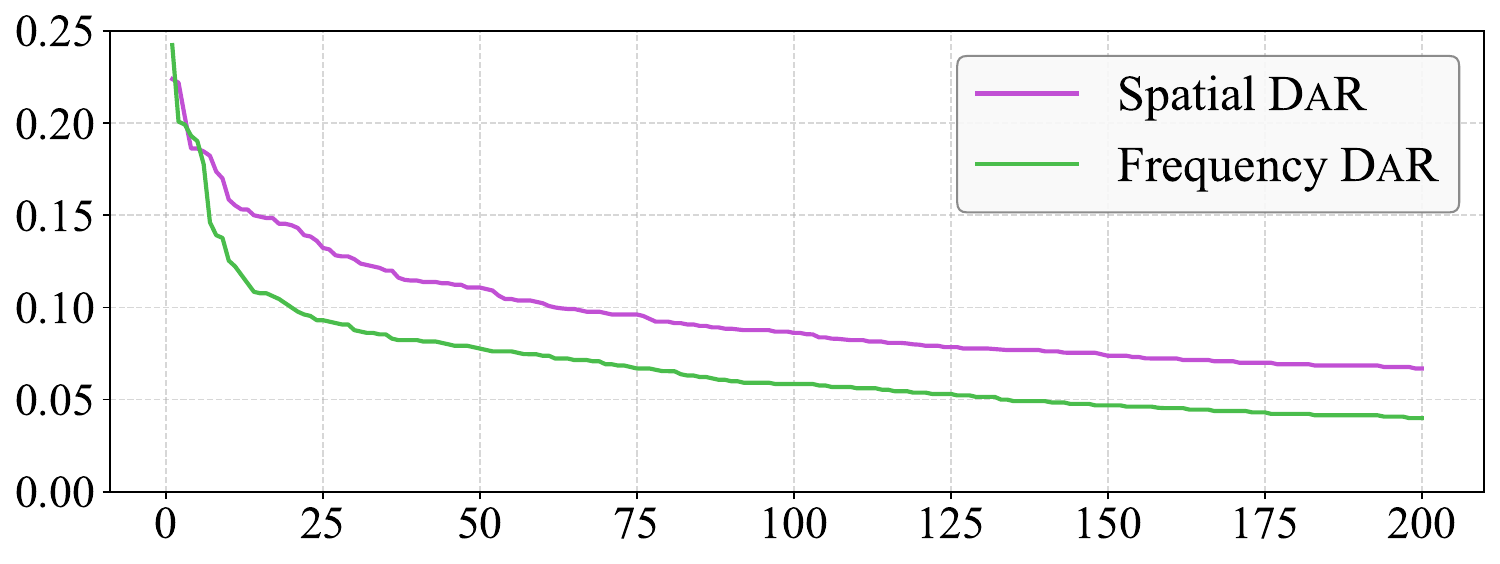}
    \caption{\textbf{Rule support is class-structured.} On ImageNet, Frequency \algname{} concentrates rule-positive samples more strongly than Spatial \algname.}
    \label{fig:long_ab}
\end{figure}

A natural question is why Frequency \algname{} appears more persistent than the spatial variant, both under clean-data distillation and under several defenses. We interpret this difference through the class-wise structure of the discovered natural triggers. \Cref{fig:long_ab} shows that spatial rules are spread more evenly across classes, whereas Frequency \algname{} produces a much more class-concentrated rule-positive distribution. This means that the frequency rule is not just an arbitrary perturbation direction or spurious noise pattern. Instead, it aligns with image statistics that are unevenly tied to particular semantic classes, making the rule closer to a task-relevant feature than an isolated artifact.

This interpretation explains the unlearning dilemma. During clean-data distillation, the student is trained to imitate the teacher on ordinary images. If the frequency rule is entangled with class-relevant clean features, clean queries repeatedly expose the same statistical relation that supports the backdoor. The student therefore has no clear signal that this relation should be discarded as malicious; it can be inherited as part of the teacher's legitimate decision rule. Conversely, a defense that tries to remove the frequency-supported behavior must suppress features that also help classify clean images from the affected classes. The defense results in \Cref{tab:backdoor_comparison} are consistent with this tension: Frequency \algname{} often retains high ASR after standard mitigation methods, while reductions in ASR are not uniformly achieved without clean-accuracy degradation or failure under other defenses. Thus, the dilemma is not simply that the backdoor is hard to erase, but that the evidence in \Cref{fig:long_ab} suggests why erasing it can conflict with preserving useful class-correlated structure.

\section{Additional OOD Support Measurements}
\label{sec:appendix_ood_support}
The main paper summarizes OOD support in \Cref{tab:ood_distillation}; here, ImageNet-C measurements show that the discovered rules remain numerically present under common corruptions and shifted image statistics.
\begin{table}[H]
    \centering
    \scriptsize
    \setlength{\tabcolsep}{3pt}
    \caption{\textbf{Rule support survives common corruptions.} On ImageNet-C, weighted trigger ratios remain close to the source poisoning rates.}
    \begin{tabular*}{\linewidth}{l @{\extracolsep{8pt}} c c @{\extracolsep{\fill}} l @{\extracolsep{8pt}} c c}
        \toprule
        Corruption & Spatial & Freq. & Corruption & Spatial & Freq. \\
        \midrule
        brightness & 2.16 & 0.48 & jpeg compression & 6.69 & 2.48 \\
        contrast & 11.49 & 25.82 & motion blur & 10.95 & 3.88 \\
        defocus blur & 11.77 & 3.68 & pixelate & 7.98 & 2.86 \\
        elastic transform & 7.30 & 2.74 & saturate & 3.86 & 1.97 \\
        fog & 13.90 & 32.37 & shot noise & 2.60 & 2.98 \\
        frost & 0.46 & 1.66 & snow & 1.46 & 0.23 \\
        gaussian blur & 11.71 & 3.85 & spatter & 6.31 & 1.57 \\
        gaussian noise & 2.47 & 3.51 & speckle noise & 4.89 & 2.85 \\
        glass blur & 10.51 & 3.80 & zoom blur & 9.03 & 4.16 \\
        impulse noise & 1.69 & 5.18 & \multicolumn{3}{c}{-} \\
        \midrule
        \multicolumn{6}{c}{\textit{Weighted Average (Overall)}} \\
        \multicolumn{3}{c}{Spatial (5\%): \textbf{6.70}} & \multicolumn{3}{c}{Frequency (3\%): \textbf{5.58}} \\
        \bottomrule
    \end{tabular*}
    \label{tab:domainshift_tight}
\end{table}

\section{Additional Design Checks}
\label{sec:appendix_design_checks}
We include two compact checks orthogonal to the main distillation result: replacing the default L2/MSE rule distance with L1 preserves high attack success, and grayscale evaluation keeps both \algname{} variants effective, indicating that the discovered relations are not restricted to color-only cues.
\begin{table}[h]
    \centering
    \caption{\textbf{The rule is not tied to one distance or color cue.} On ImageNet/ResNet18, \algname{} remains effective with L1 distance and grayscale evaluation.}
    \label{tab:additional_design_checks}
    \footnotesize
    \begin{subtable}[t]{0.46\linewidth}
        \centering
        \caption{L1 rule distance}
        \setlength{\tabcolsep}{5pt}
        \begin{tabular*}{\linewidth}{@{\extracolsep{\fill}}lcc}
            \toprule
            Method & CA & ASR \\
            \midrule
            No Attack & 69.76 & - \\
            Spatial \algname{} & 67.28 & 99.97 \\
            Frequency \algname{} & 69.02 & 95.08 \\
            \bottomrule
        \end{tabular*}
    \end{subtable}\hfill
    \begin{subtable}[t]{0.46\linewidth}
        \centering
        \caption{Grayscale evaluation}
        \setlength{\tabcolsep}{5pt}
        \begin{tabular*}{\linewidth}{@{\extracolsep{\fill}}lcc}
            \toprule
            Method & CA & ASR \\
            \midrule
            No Attack & 68.27 & - \\
            Spatial \algname{} & 65.93 & 99.77 \\
            Frequency \algname{} & 66.28 & 99.96 \\
            \bottomrule
        \end{tabular*}
    \end{subtable}
\end{table}

\section{Types of Attacks}
\label{sec:attack_types}
In this section, we provide a comprehensive overview of backdoor attack taxonomies and justify our experimental setup that uses four representative attack methods: BadNets~\cite{badnets}, Blend~\cite{blend}, Refool~\cite{refool}, and WaNet~\cite{wanet}. These methods cover the major backdoor attack types used in our evaluation.
\section{Qualitative Results}
\label{sec:qualitative}
To provide a concrete understanding of our method's operation, Table~\ref{tab:poisoned_examples} presents qualitative examples of the poisoning process. Specifically, the left column displays images selected as targeted poison samples according to our criteria, while the right column shows the results after injecting triggers into benign images. These examples illustrate the selection strategy for targeted poisoning and the visual characteristics of poisoned samples generated by our attack method in the appendix.
\begin{table*}[h]
    \centering
    \scriptsize
    \caption{\textbf{Selected clean positives define the test-time rule.} ImageNet examples pair mined Spatial/Frequency positives with their triggered variants.}
    \label{tab:poisoned_examples}
    \setlength{\tabcolsep}{2pt}
    \begin{tabular}{l|c|cc|ccc}
        \toprule
        & \multicolumn{6}{c}{Poisoned samples} \\
        \cmidrule(lr){2-7}
        Description &
        Ex.1 & Ex.2 & Ex.3 & Ex.4 & Ex.5 & Ex.6 \\
        \midrule
        Spatial \algname{} &
        \includegraphics[width=0.1\textwidth]{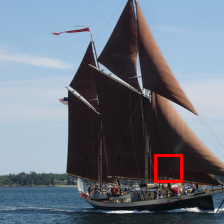} &
        \includegraphics[width=0.1\textwidth]{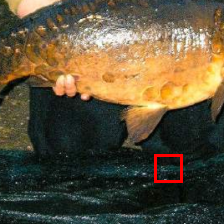} &
        \includegraphics[width=0.1\textwidth]{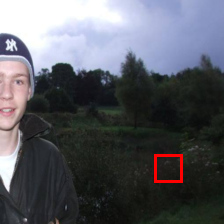} &
        \includegraphics[width=0.1\textwidth]{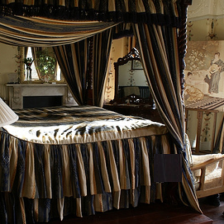} &
        \includegraphics[width=0.1\textwidth]{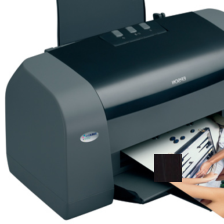} &
        \includegraphics[width=0.1\textwidth]{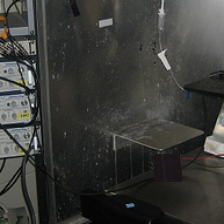} \\
        Frequency \algname&
        \includegraphics[width=0.1\textwidth]{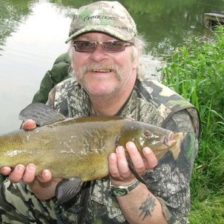} &
        \includegraphics[width=0.1\textwidth]{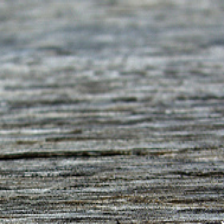} &
        \includegraphics[width=0.1\textwidth]{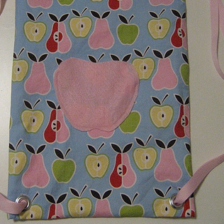} &
        \includegraphics[width=0.1\textwidth]{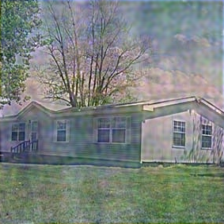} &
        \includegraphics[width=0.1\textwidth]{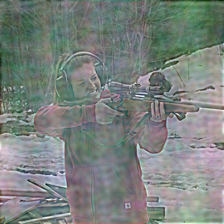} &
        \includegraphics[width=0.1\textwidth]{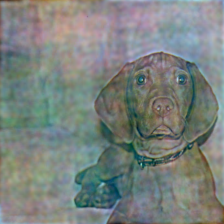} \\
        \bottomrule
    \end{tabular}
    \label{tab:qualitative}
\end{table*}

\subsection{Attack Taxonomy}
\label{subsec:taxonomy}
Backdoor attacks can be categorized by trigger type and its relationship to the input data. Based on these characteristics, we classify attacks into static and dynamic attacks for the evaluation taxonomy.
\subsection{Static Attacks}
\label{subsec:static}
Static attacks use consistent triggers and can be subdivided by spatial characteristics:
\paragraph{Local Trigger Attacks.}
\textbf{Representative method: BadNets~\cite{badnets}}
Local trigger attacks embed a fixed patch or symbol at a chosen position. Key characteristics include consistent spatial placement:
\begin{itemize}
    \item Fixed position, typically in a corner or less salient region of the image
    \item Consistent pattern across all poisoned samples
    \item Relatively easy to implement but also easier to detect
    \item Direct modification of pixel values in a localized region
\end{itemize}
BadNets exemplifies this approach by inserting a small fixed pattern, such as a square, at a consistent location. It is highly effective but leaves a distinctive signature identifiable by stronger defenses.
\paragraph{Global Trigger Attacks.}
\textbf{Representative method: Blend~\cite{blend}}
Global trigger attacks distribute the trigger pattern across all or most of the input image. Key characteristics include:
\begin{itemize}
    \item Pattern spans the entire input space rather than a localized region
    \item Often implemented through blending or transparency effects
    \item More subtle visual appearance compared to local triggers
    \item May be less disruptive to semantic content
\end{itemize}
Blend superimposes a controlled-transparency pattern over the entire image, making it less visually obvious than patch attacks while preserving high attack success in practice across poisoned inputs.
\paragraph{Multiple Trigger Attacks.}
\textbf{Representative method: Refool~\cite{refool}}
Multiple trigger attacks employ several different triggers that all map to the same target label. Their key characteristics include:
\begin{itemize}
    \item Uses multiple distinct trigger patterns
    \item All triggers direct classification to the same target
    \item Increases attack robustness against defenses that may only identify some triggers
    \item Often leverages naturally occurring visual elements
\end{itemize}
Refool uses reflection patterns as triggers, embedding natural-looking reflection effects that strongly influence predictions and remain robust against many defenses and detectors.
\subsection{Dynamic Attacks}
\label{subsec:dynamic}
\textbf{Representative method: WaNet~\cite{wanet}}
Unlike static attacks, dynamic attacks use input-dependent triggers that adapt to each image and its content. Their key characteristics include:
\begin{itemize}
    \item Trigger transformation depends on the input image content
    \item No fixed pattern that repeats across poisoned samples
    \item Often implemented through subtle warping or transformations
    \item Significantly harder to detect through statistical analysis
\end{itemize}
WaNet implements this approach through imperceptible image warping, where the warping function creates an image-specific trigger. This makes the attack difficult to defend against because traditional methods that search for consistent poisoned-input patterns fail to identify the dynamic trigger.

\section{Defense Evaluation Details}
\label{sec:defense_details}
We separate the defenses in \Cref{tab:backdoor_comparison} into two groups because they answer different operational questions. Training/model-level mitigation methods modify the model and are evaluated by the resulting clean accuracy (CA) and attack success rate (ASR). Detection methods leave the model fixed and instead reject suspicious inputs; for these methods, we report the clean block rate and the effective ASR after rejected attack inputs are removed.

\paragraph{Training and model-level mitigation.}
FT fine-tunes the backdoored model on a small clean subset. FP prunes low-activation channels and then fine-tunes the remaining model. SAU and NAD use clean data to unlearn the backdoor behavior through adversarial unlearning or attention distillation. CLP removes channels using channel Lipschitzness scores without relying on poisoned samples. OTBR fuses the defended model with a pruned counterpart through optimal transport. Unless a method is data-free by design, it is given 1\% of ImageNet training data. For CLP, we sweep the pruning threshold and report the best valid point under a clean-accuracy constraint, avoiding degenerate settings that erase ASR only by collapsing clean accuracy.

\paragraph{Detection defenses.}
STRIP detects inputs with low prediction entropy under input superposition. Spectral scores samples through representation-space anomalies. We calibrate detection thresholds on clean validation samples at a target false positive rate of 5\%. The table reports clean block rate and effective ASR after detection, so an effective detector should keep the clean block rate low while sharply reducing ASR. This reporting is distinct from CA/ASR after mitigation because detectors reject inputs rather than repair classifiers during inference-time filtering and deployment.

\end{document}